\documentclass[aps,prl,twocolumn,showpacs,superscriptaddress,groupedaddress]{revtex4}  % for review and submission
\usepackage{graphicx}  % needed for figures
\usepackage{dcolumn}   % needed for some tables
\usepackage{bm}        % for math
\usepackage{amssymb}   % for math
\usepackage{color}     % for color

\def\lsim{\mathrel{\rlap{\lower4pt\hbox{\hskip1pt$\sim$}}\raise1pt\hbox{$<$}}}
\def\gsim{\mathrel{\rlap{\lower4pt\hbox{\hskip1pt$\sim$}}\raise1pt\hbox{$>$}}}
\def\MET{{\mbox{$E\kern-0.57em\raise0.19ex\hbox{/}_{T}~$}}}
\def\METnoSpace{{\mbox{$E\kern-0.57em\raise0.19ex\hbox{/}_{T}$}}}
\def\dzero{D0}
\def\ww{$WW$}
\def\wz{$WZ$}
\def\gww{{\mbox{$G\rightarrow WW$}}}

\begin{document}
%\setpagewiselinenumbers
%\modulolinenumbers[5]
%\linenumbers

% the following line is for submission, including submission to the arXiv!!
\hspace{5.2in} \mbox{FERMILAB-PUB-10/485-E}

\title{Search for resonant $\boldsymbol{WW}$~and $\boldsymbol{WZ}$~production in
$\boldsymbol{p\bar{p}}$~collisions at $\boldsymbol{\sqrt{s} = 1.96}$~TeV}
% remove these 3 lines before journal submittal.
%\centerline{author list dated 6 October 2010}
% end removal before journal submittal
%
\affiliation{Universidad de Buenos Aires, Buenos Aires, Argentina}
\affiliation{LAFEX, Centro Brasileiro de Pesquisas F{\'\i}sicas, Rio de Janeiro, Brazil}
\affiliation{Universidade do Estado do Rio de Janeiro, Rio de Janeiro, Brazil}
\affiliation{Universidade Federal do ABC, Santo Andr\'e, Brazil}
\affiliation{Instituto de F\'{\i}sica Te\'orica, Universidade Estadual Paulista, S\~ao Paulo, Brazil}
\affiliation{Simon Fraser University, Vancouver, British Columbia, and York University, Toronto, Ontario, Canada}
\affiliation{University of Science and Technology of China, Hefei, People's Republic of China}
\affiliation{Universidad de los Andes, Bogot\'{a}, Colombia}
\affiliation{Charles University, Faculty of Mathematics and Physics, Center for Particle Physics, Prague, Czech Republic}
\affiliation{Czech Technical University in Prague, Prague, Czech Republic}
\affiliation{Center for Particle Physics, Institute of Physics, Academy of Sciences of the Czech Republic, Prague, Czech Republic}
\affiliation{Universidad San Francisco de Quito, Quito, Ecuador}
\affiliation{LPC, Universit\'e Blaise Pascal, CNRS/IN2P3, Clermont, France}
\affiliation{LPSC, Universit\'e Joseph Fourier Grenoble 1, CNRS/IN2P3, Institut National Polytechnique de Grenoble, Grenoble, France}
\affiliation{CPPM, Aix-Marseille Universit\'e, CNRS/IN2P3, Marseille, France}
\affiliation{LAL, Universit\'e Paris-Sud, CNRS/IN2P3, Orsay, France}
\affiliation{LPNHE, Universit\'es Paris VI and VII, CNRS/IN2P3, Paris, France}
\affiliation{CEA, Irfu, SPP, Saclay, France}
\affiliation{IPHC, Universit\'e de Strasbourg, CNRS/IN2P3, Strasbourg, France}
\affiliation{IPNL, Universit\'e Lyon 1, CNRS/IN2P3, Villeurbanne, France and Universit\'e de Lyon, Lyon, France}
\affiliation{III. Physikalisches Institut A, RWTH Aachen University, Aachen, Germany}
\affiliation{Physikalisches Institut, Universit{\"a}t Freiburg, Freiburg, Germany}
\affiliation{II. Physikalisches Institut, Georg-August-Universit{\"a}t G\"ottingen, G\"ottingen, Germany}
\affiliation{Institut f{\"u}r Physik, Universit{\"a}t Mainz, Mainz, Germany}
\affiliation{Ludwig-Maximilians-Universit{\"a}t M{\"u}nchen, M{\"u}nchen, Germany}
\affiliation{Fachbereich Physik, Bergische  Universit{\"a}t Wuppertal, Wuppertal, Germany}
\affiliation{Panjab University, Chandigarh, India}
\affiliation{Delhi University, Delhi, India}
\affiliation{Tata Institute of Fundamental Research, Mumbai, India}
\affiliation{University College Dublin, Dublin, Ireland}
\affiliation{Korea Detector Laboratory, Korea University, Seoul, Korea}
\affiliation{CINVESTAV, Mexico City, Mexico}
\affiliation{FOM-Institute NIKHEF and University of Amsterdam/NIKHEF, Amsterdam, The Netherlands}
\affiliation{Radboud University Nijmegen/NIKHEF, Nijmegen, The Netherlands}
\affiliation{Joint Institute for Nuclear Research, Dubna, Russia}
\affiliation{Institute for Theoretical and Experimental Physics, Moscow, Russia}
\affiliation{Moscow State University, Moscow, Russia}
\affiliation{Institute for High Energy Physics, Protvino, Russia}
\affiliation{Petersburg Nuclear Physics Institute, St. Petersburg, Russia}
\affiliation{Stockholm University, Stockholm and Uppsala University, Uppsala, Sweden }
\affiliation{Lancaster University, Lancaster LA1 4YB, United Kingdom}
\affiliation{Imperial College London, London SW7 2AZ, United Kingdom}
\affiliation{The University of Manchester, Manchester M13 9PL, United Kingdom}
\affiliation{University of Arizona, Tucson, Arizona 85721, USA}
\affiliation{University of California Riverside, Riverside, California 92521, USA}
\affiliation{Florida State University, Tallahassee, Florida 32306, USA}
\affiliation{Fermi National Accelerator Laboratory, Batavia, Illinois 60510, USA}
\affiliation{University of Illinois at Chicago, Chicago, Illinois 60607, USA}
\affiliation{Northern Illinois University, DeKalb, Illinois 60115, USA}
\affiliation{Northwestern University, Evanston, Illinois 60208, USA}
\affiliation{Indiana University, Bloomington, Indiana 47405, USA}
\affiliation{Purdue University Calumet, Hammond, Indiana 46323, USA}
\affiliation{University of Notre Dame, Notre Dame, Indiana 46556, USA}
\affiliation{Iowa State University, Ames, Iowa 50011, USA}
\affiliation{University of Kansas, Lawrence, Kansas 66045, USA}
\affiliation{Kansas State University, Manhattan, Kansas 66506, USA}
\affiliation{Louisiana Tech University, Ruston, Louisiana 71272, USA}
\affiliation{Boston University, Boston, Massachusetts 02215, USA}
\affiliation{Northeastern University, Boston, Massachusetts 02115, USA}
\affiliation{University of Michigan, Ann Arbor, Michigan 48109, USA}
\affiliation{Michigan State University, East Lansing, Michigan 48824, USA}
\affiliation{University of Mississippi, University, Mississippi 38677, USA}
\affiliation{University of Nebraska, Lincoln, Nebraska 68588, USA}
\affiliation{Rutgers University, Piscataway, New Jersey 08855, USA}
\affiliation{Princeton University, Princeton, New Jersey 08544, USA}
\affiliation{State University of New York, Buffalo, New York 14260, USA}
\affiliation{Columbia University, New York, New York 10027, USA}
\affiliation{University of Rochester, Rochester, New York 14627, USA}
\affiliation{State University of New York, Stony Brook, New York 11794, USA}
\affiliation{Brookhaven National Laboratory, Upton, New York 11973, USA}
\affiliation{Langston University, Langston, Oklahoma 73050, USA}
\affiliation{University of Oklahoma, Norman, Oklahoma 73019, USA}
\affiliation{Oklahoma State University, Stillwater, Oklahoma 74078, USA}
\affiliation{Brown University, Providence, Rhode Island 02912, USA}
\affiliation{University of Texas, Arlington, Texas 76019, USA}
\affiliation{Southern Methodist University, Dallas, Texas 75275, USA}
\affiliation{Rice University, Houston, Texas 77005, USA}
\affiliation{University of Virginia, Charlottesville, Virginia 22901, USA}
\affiliation{University of Washington, Seattle, Washington 98195, USA}
\author{V.M.~Abazov} \affiliation{Joint Institute for Nuclear Research, Dubna, Russia}
\author{B.~Abbott} \affiliation{University of Oklahoma, Norman, Oklahoma 73019, USA}
\author{B.S.~Acharya} \affiliation{Tata Institute of Fundamental Research, Mumbai, India}
\author{M.~Adams} \affiliation{University of Illinois at Chicago, Chicago, Illinois 60607, USA}
\author{T.~Adams} \affiliation{Florida State University, Tallahassee, Florida 32306, USA}
\author{G.D.~Alexeev} \affiliation{Joint Institute for Nuclear Research, Dubna, Russia}
\author{G.~Alkhazov} \affiliation{Petersburg Nuclear Physics Institute, St. Petersburg, Russia}
\author{A.~Alton$^{a}$} \affiliation{University of Michigan, Ann Arbor, Michigan 48109, USA}
\author{G.~Alverson} \affiliation{Northeastern University, Boston, Massachusetts 02115, USA}
\author{G.A.~Alves} \affiliation{LAFEX, Centro Brasileiro de Pesquisas F{\'\i}sicas, Rio de Janeiro, Brazil}
% ...
%
% visitor_addresses.tex                        6 October 2010
%  available symbols are:
%  $\ast, \dag, \ddag, \S, \P, $\|$, $\ast\ast$, \dag\dag, \ddag\ddag ,\#
%
\collaboration{The D0 Collaboration\footnote{with visitors from
%{alton}
$^{a}$Augustana College, Sioux Falls, SD, USA,
%{burdin}
$^{b}$The University of Liverpool, Liverpool, UK,
%{haas,partridge}
$^{c}$SLAC, Menlo Park, CA, USA,
%{juste}
$^{d}$ICREA/IFAE, Barcelona, Spain,
%{luna-garcia}
$^{e}$Centro de Investigacion en Computacion - IPN, Mexico City, Mexico,
%{podesta-lerma}
$^{f}$ECFM, Universidad Autonoma de Sinaloa, Culiac\'an, Mexico,
and 
%{weber}
$^{g}$Universit{\"a}t Bern, Bern, Switzerland.%
%{hooper}
%$^{?}$%Visitor from Bradley University, Peoria, IL, USA.
%{kozminski
%$^{?}$}%Visitor from Lewis University, Romeoville, IL, USA.
%{deceased}
%$^{\ddag}$%Deceased.
}} \noaffiliation
\vskip 0.25cm
       % D0 authors (remove the first 3 lines
                             % of this file prior to submission, they
                             % contain a time stamp for the authorlist)
                             % (includes institutions and visitors)
\date{November 29, 2010}

\begin{abstract}
We search for resonant \ww~or \wz~production using up to $5.4$~fb$^{-1}$ of
integrated luminosity collected by the \dzero~experiment in Run II of the
Fermilab Tevatron Collider. The data are consistent with the standard model background expectation, and we set limits on a
resonance mass using the sequential standard model (SSM) $W'$~boson and the Randall-Sundrum
model graviton $G$~as benchmarks.  We exclude an SSM $W'$~boson in the mass range
$180-690$~GeV and a Randall-Sundrum graviton in the range $300-754$~GeV at $95\%$~CL.
\end{abstract}

\maketitle

% INTRODUCTION & MOTIVATION
%%%%%%%%%%%%%%%%%%%%%%%%%%%%%%%%%%%%%%%%%%%%%%%%%%%%%%%%%%%%%%%%%%%%%%%%%%%
The standard model of particle physics is expected to be a low energy
effective theory valid for particle interactions below the TeV scale. Above this
scale, extensions to the standard model (SM) augment the existing particle
content, leading to enhanced production of many
final states at colliders. Specifically, the production and decay of massive charged or
neutral particles can produce an excess of $W$~boson pairs for neutral particles
or $W$~and $Z$~boson pairs for charged particles. 

In this Letter, we search for resonant \ww~and~\wz~production using data
collected by the \dzero~detector from 1.96~TeV $p\bar{p}$~collisions produced by
the Fermilab Tevatron Collider. We use a sequential standard model (SSM)~\cite{ssm,large_width} $W'$~boson
as benchmark for a $WZ$~resonance and a Randall-Sundrum
(RS)~\cite{rs1,rs2,rs3}~graviton (G) resonance for the \ww~final state.

There are two recent direct searches for \wz~or~\ww~resonances by the CDF and
\dzero~collaborations that exclude~\wz~resonances with mass below 516 and 520~GeV,
respectively, and an RS graviton \gww~resonance with mass less
than 607~GeV~\cite{CDFrun2vv,wzlvll}. Indirect searches for new physics in the
\ww~and \wz~diboson systems through measurements of the triple gauge couplings also 
show no deviation from the SM predictions~\cite{aTGC1,aTGC2,aTGC3}. The
\dzero~collaboration also excludes $M(W')<1.00$~TeV~\cite{wprime_ev_d0}, when
assuming the $W'$~boson decays as in the SM, and $M(G)<1.05$~TeV~\cite{graviton_d0} when assuming $G$~decays into $\gamma\gamma$~or
$ee$. The search for these resonances in the diboson decay channel
covers the possibility that their coupling to leptons may
be lower than the value predicted by the SM.

% DATA
%%%%%%%%%%%%%%%%%%
We obtain a combined result based on three
independent searches: two new searches for resonant $WW/WZ$~production with at least one jet and exactly
one or two leptons in the final state using $5.4$~fb$^{-1}$~of integrated
luminosity and one search previously done on $4.1$~fb$^{-1}$ of integrated luminosity with three
leptons in the final state~\cite{wzlvll}. We use data collected by the
\dzero~experiment. A detailed description of the \dzero~detector can be found
in~\cite{d0det}; we only give a brief description here. The
innermost region is the tracking detector, which consists of silicon microstrip and central fiber trackers, both of which are
surrounded by a solenoidal magnet producing a 2 T magnetic field. Charged
particle tracks are formed from signals in these detectors. Surrounding the tracking
detector are electromagnetic (EM) and hadronic calorimeters, both of which use liquid argon as
the active medium. The calorimeters are housed in three cryostats that define the
central region as $|\eta|<1.1$~\cite{deltar}~and two endcap regions as
$1.5<|\eta|<4$. Electrons are reconstructed in the EM calorimeter as
isolated energy clusters, matched to tracks, and with a shower shape that is consistent
with that of an electron. Jets are also formed in the
calorimeters as clusters of energy in a cone with radius ${\cal R}=0.5$~\cite{d0jets}. Finally,
surrounding the
calorimeters are central and forward muon systems in three layers consisting of precision
wire chambers and fast scintillators used for triggering. Coverage
of the muon system extends to $|\eta|\approx2$. Located between the first and
second layer of the muon system is a 1.8 T toroidal magnet which allows an
independent muon momentum measurement. A muon candidate is reconstructed as the
combination of tracks in the muon system and the inner tracking detector and is
required to be isolated from other tracks or calorimeter energy deposits.

% SAMPLES
%%%%%%%%%%%%%%%%%%
We employ a Monte Carlo (MC) simulation to model all background processes except
backgrounds from events not involving the decay of a $W$~or $Z$~boson, such as multijet production. The SSM $W'$~boson, RS graviton, and the SM diboson processes are simulated using
the {\sc pythia}~\cite{pythia}~event generator, which generates the tree-level
matrix element process and simulates subsequent particle showering and hadronization
effects. Backgrounds from $t\bar{t}$, $W$+jets with $W\rightarrow\ell\nu$, and
$Z$+jets with $Z\rightarrow\ell\ell$~are modeled using the {\sc alpgen}~\cite{alpgen}
generator and single top quark production is modeled with the {\sc comphep}~\cite{comphep}~generator.
All generators are interfaced with {\sc pythia} for showering and
hadronization. The {\sc alpgen}-generated samples make use of the MLM~\cite{mlm} jet-parton
matching scheme to improve the jet multiplicity modeling. All MC samples are passed through a {\sc
geant}-based~\cite{geant} simulation of the D0 detector and overlayed with data events
from randomly selected bunch crossings to simulate multiple
$p\bar{p}$~interactions within a single event. The signal samples are generated
in exclusive final states with diboson resonance masses between 180 and 1250~GeV
in 10~GeV steps up to 200~GeV and then 50~GeV steps above 200~GeV, using the CTEQ6L1~\cite{cteq} parton
distribution functions. No interference between the SM $W$~boson and the SSM
$W'$~boson production is included in the simulation since
the effect is negligible~\cite{rizzo}. All MC samples are
normalized such that the predicted yield is equal to the production cross
section multiplied by the integrated luminosity of data. The $W$+jets and $Z$+jets samples are scaled to
the product of the cross section calculated by {\sc alpgen} and the
$k$-factor defined as the ratio of the next-to-leading order (NLO) and leading order
(LO) cross sections, which is computed
by {\sc MCFM}~\cite{MCFM}. The $t\bar{t}$~events are normalized to a
next-to-next-to-leading order (NNLO) calculation~\cite{moch:uwer} with
$m_{\rm{t}}=172.5$~GeV. Finally, the diboson samples are normalized to the NLO cross section
predicted by {\sc MCFM}, and the signal $W'$~boson samples are normalized to the NNLO
cross sections~\cite{wprime_nnlo}. The RS graviton samples are normalized to the
{\sc pythia}-level cross section multiplied by a $k$-factor of 1.3~\cite{graviton_kf}.

% OBJECT SELECTION
%%%%%%%%%%%%%%%%%%
Events in this search are placed in three mutually exclusive categories,
thus maximizing signal sensitivity to each \ww~and \wz~decay channel. The
first category contains events with a leptonic decay of the $W$~boson and
hadronic decay of the $W$~or $Z$~boson. Events must contain exactly one electron or muon
with transverse momentum $p_{T}>20$~GeV, either one or two jets with
$p_{T}>20$~GeV, and missing transverse energy
$\METnoSpace>20$~GeV~\cite{metdef}. These events were collected using triggers
that require the presence of a high $p_{T}$~lepton. Events with charged leptonic decays of the $Z$~boson and hadronic decays of the
$W$~boson comprise the second category. These events must contain exactly two electrons or muons and exactly one or two jets with the same
$p_{T}$~thresholds as the first category. We require $\METnoSpace<50$~GeV to remove mismeasured events and a
dilepton mass between 70 and 110~GeV to select $Z$~boson events. Both single
lepton and dilepton triggers were used to collect events in this category. Fully leptonic decays of the $WZ$~system constitute the final
selection category. In this category any combination of three leptons ($eee$,
 $\mu\mu\mu$, $e\mu\mu$, $ee\mu$) with $p_{T}>20$~GeV for each lepton is
accepted. Additionally, $\METnoSpace>30$~GeV is required. Events
in this channel were collected using the same set of triggers as the dilepton
channel. More details of the trilepton analysis
have been presented in a previous Letter~\cite{wzlvll}.

% BACKGROUNDS
%%%%%%%%%%%%%%%%%%
In the first two selection categories the background after the initial event selection is dominated by
$W$~or $Z$~boson+jets, followed by multijet, $t\bar{t}$, single top quark, and
diboson production. The multijet background, in both single lepton and dilepton events, is
modeled using data that fail the final lepton quality selection criteria. In single lepton
events, the relative fraction of multijet
({\it fake}-lepton) background and all other backgrounds ({\it real}-lepton) is
determined by measuring the relative rates at which each background type
satisfies two different lepton quality criteria.  A sample of
$Z\rightarrow\ell\ell$~events is used to measure the {\it real}-lepton efficiency, and
a sample of events without significant missing transverse energy
($\METnoSpace<20$~GeV) is used to measure the {\it fake}-lepton efficiency. In
dilepton events, the sum of the MC-based backgrounds and the multijet background
is normalized to data in the dilepton mass region between 40 and 70 GeV. The
signal acceptance, in both single lepton and dilepton events, is estimated
from the MC and corrected for all data/MC differences and the estimated trigger selection efficiency.

% Signal selection
%%%%%%%%%%%%%%%%%%%
The $W$~and $Z$~bosons result from the decay of a massive resonance and are therefore highly boosted. We exploit this
property in the single lepton channel by requiring that the $p_{T}$~of the
lepton-$\METnoSpace$~system be greater than 100~GeV
and the azimuthal angle between the lepton and $\METnoSpace$ be less than
1.5~radians. In the dilepton channel, we require that the dilepton pair $p_{T}$~be
greater than 100 GeV. The angular distance $\Delta{\cal R}$~between the two leptons
must be less than 1.5, where the angular distance $\Delta{\cal R}=\sqrt{(\Delta\eta)^{2}+(\Delta\phi)^{2}}$~is defined as the
distance between the two objects in the
pseudorapidity and azimuthal angle plane. The hadronically decaying
$W$~or $Z$~boson in both single lepton and dilepton events will also be highly
boosted. Given the extended
size of jets, the two jets from the hadronic decay of $W$~or $Z$~bosons with
sufficient transverse momentum may be merged in a single jet whose mass, defined
in terms of the jet energy $E_{j}$~and momentum $\vec{p}_{j}$~as $m_{j}=\sqrt{E_{j}^{2}-|\vec{p_{j}}|^{2}}$, corresponds to the original boson mass. Thus, we further
select events having a single jet with $m_{j}$~greater than $60(70)$~GeV
when searching for a hadronically decaying $W(Z)$~boson. If no jets satisfy these requirements, two isolated jets must be
reconstructed in the event with a dijet mass between 60 and 105~GeV~(70 and
115~GeV) for a $W(Z)$~boson decay. In each channel, the $\Delta{\cal R}$~between the jets must be
less than 1.5 radians.

\begin{figure}
\includegraphics[scale=0.45]
{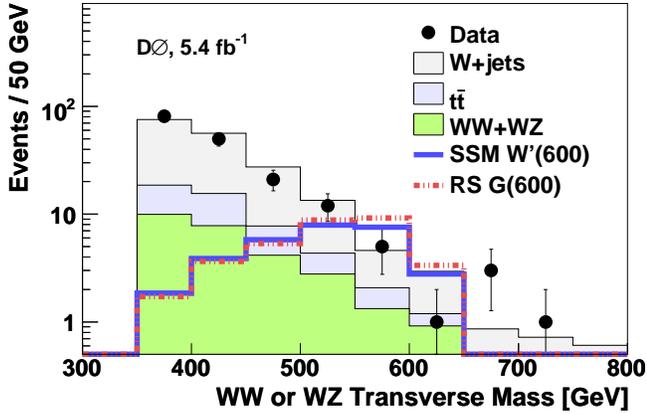}
\caption{\label{fig:plot1} (color online) The reconstructed $WW$~or~$WZ$~transverse mass distribution for the data
(points) with statistical error bars and the estimated $W$+jets, $t\bar{t}$, and diboson backgrounds in the single
lepton channel. The predicted SSM $W'$~distribution for $M(W')=600$~GeV and the RS
graviton distribution for $M(G)=600$~GeV are shown as solid and dashed lines,
respectively.}
\end{figure}

\begin{figure}
\includegraphics[scale=0.45]
{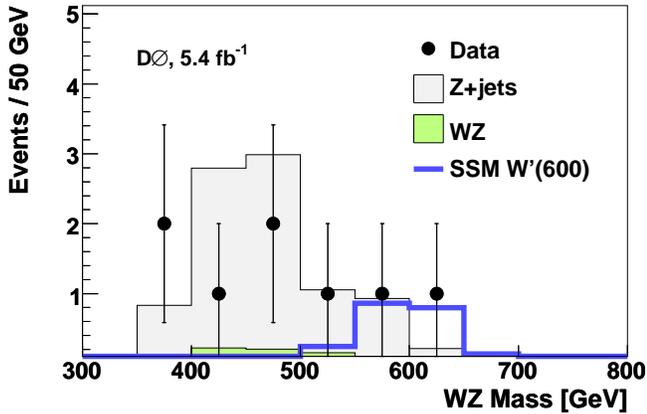}
\caption{\label{fig:plot2} (color online) The reconstructed $WZ$~mass distribution for the data
(points) with statistical error bars and the estimated $Z$+jets and diboson backgrounds, and the predicted
SSM $W'$~distribution for $M(W')=600$~GeV in the dilepton channel.}
\end{figure}

We increase the search sensitivity by subdividing all search channels into
``high''-mass and ``low''-mass signal regions, where the
mass refers to the assumed signal mass ($M_{\rm{res}}$) and high-mass and
low-mass are defined as $M_{\rm{res}}\geq450$~GeV~and~$<450$~GeV,
respectively. The low-mass signal region is composed of all events that satisfy the
single lepton and dilepton selection requirements. In the high-mass signal region we additionally require the lepton-$\METnoSpace$~system~$p_{T}>150$~GeV and the azimuthal angle
difference between the lepton and $\METnoSpace$~to be less than 1.0 radian for single lepton
events. The dilepton high-mass selection requires the dilepton $p_{T}$~to be greater than $150$~GeV
and the $\Delta{\cal R}$~between the two leptons must to be less than 1.0 radian. Table~\ref{tab:table1}~displays the estimated background yields, the
expected numbers of signal events, and the numbers of observed data events after
the high-mass selection. Figs.~\ref{fig:plot1}~and~\ref{fig:plot2}
compare the data with the estimated backgrounds in the single lepton and
dilepton channels using the reconstructed resonance mass and
transverse mass~\cite{mtwz}, respectively. In both Figures, the absence of
events below 350~GeV is a result of the high-mass event selection.

The dominant systematic uncertainties on the background normalization and signal
acceptance in the dilepton channel are mostly due to
$Z$+jets modeling ($10\%$) and to jet energy resolution ($3\%$) effects. The main
uncertainties in single lepton events are $W$+jets modeling ($15\%$),
the $t\bar{t}$~cross section ($10\%$), and the integrated luminosity
($6.1\%$)~\cite{lumi}. In single jet events, the principal uncertainty is the
jet-mass modeling. To determine it, a control region is
defined from events satisfying the initial event selection, but failing the
lepton-$\METnoSpace$ or the dilepton system $p_{T} > 100$~GeV requirement. The relative
difference between the background prediction and the data is $10\%$~for jet
masses below $30$~GeV and rises to $25\%$~for masses near~$60$~GeV. We do not observe any event with jets having a mass
greater than 60 GeV in the control region.

\begin{table}
\caption{\label{tab:table1}Expected event yields for the high-mass single
lepton and dilepton selection samples. Uncertainties are combined statistical and systematic
uncertainties on the background yields.}
\begin{ruledtabular}
\begin{tabular}{lD{,}{\,\pm\,}{-1}D{,}{\,\pm\,}{-1}}
Process&\multicolumn{1}{c}{Single lepton sample}&\multicolumn{1}{r}{Dilepton sample}\\
\hline
$Z$+jets	&    3.6, 0.2		&   7.9, 0.8	\\
$W$+jets        &    124.5, 20.3	&   <0.01 	\\
Top             &    22.9, 2.5		&   <0.01	\\
Multijet        &    4.6, 0.3		&   <0.01	\\
Diboson         &    27.6, 1.4		&   0.8, 0.1	\\ 
\hline
Background sum  &    183.2, 24.5	&   8.7, 0.8	\\
\hline
Data            &    174  & 8 \\
\end{tabular}
\end{ruledtabular}
\end{table}

\begin{figure}
\includegraphics[scale=0.45]
{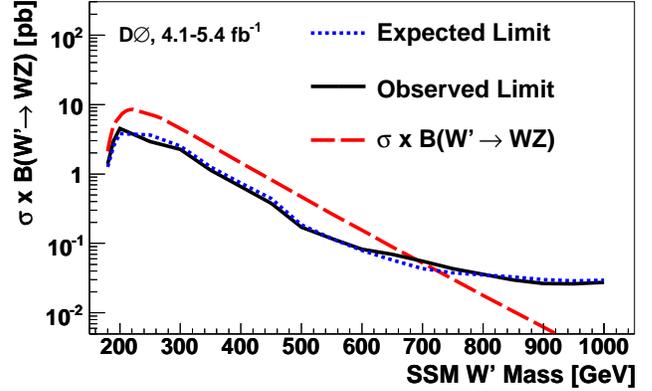}
\caption{\label{fig:plot3} (color online) The excluded production cross section for an SSM
$W'$~boson as a function of the $W'$~boson mass. The excluded mass region is the region
where the SSM $W'$~cross section exceeds the excluded cross section.}
\end{figure}

\begin{figure}
\includegraphics[scale=0.45]
{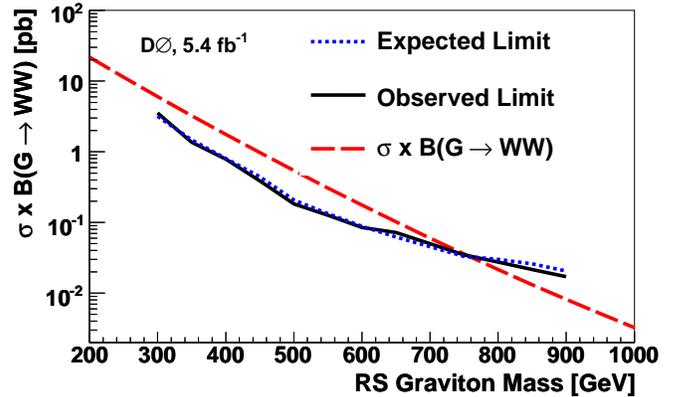}
\caption{\label{fig:plot4} (color online) The excluded production cross section for an RS
graviton as a function of the graviton mass. The excluded mass region is the region where the RS graviton cross section exceeds the excluded cross section.}
\end{figure}

%%%%%%%%%%%%%%%%%%
% LIMITS
%%%%%%%%%%%%%%%%%%
No statistically significant excess of the data over the background prediction
is observed. Thus, we set limits on the production cross section multiplied by
the branching ratio using a modified frequentist approach~\cite{collie}. In this method a log-likelihood
ratio (LLR) test statistic~\cite{llr} is formed using the Poisson probabilities for estimated background yields,
the signal acceptance, and the observed number of events for all resonant mass
hypotheses. Confidence levels are derived by integrating the LLR~in
pseudo-experiments using both the signal plus background hypotheses
($CL_{s+b}$)~as well as the
background only hypothesis ($CL_{b}$). In the modified frequentist approach,
the excluded production cross section is computed as the cross section for
which $CL_{s}$, defined as $CL_{s+b}/CL_{b}$, is equal to
0.05. Limits on the $WZ$ resonance are set using the distribution
of the reconstructed mass in dilepton events and of the reconstructed
transverse mass in single lepton events, using a bin size of 50 GeV.
The $W'$~signal acceptance and the expected background yield are
parametrized accordingly. A similar procedure is used to set limits
on the RS graviton signal using the reconstructed transverse mass
in single lepton events. We exclude an SSM~$W'$~boson in the mass region between 180 and
690~GeV and an RS graviton in the mass range of 300 to 754~GeV at $95\%$~CL. The graviton mass
exclusion assumes the dimensionless coupling parameter
$k/\overline{M}_{pl}=0.1$, where $k$~is the curvature scale of the warped extra
dimension and $\overline{M}_{pl}=M_{pl}/\sqrt{8\pi}$ is the reduced Planck mass. The expected and observed excluded cross sections as a function of the signal resonance mass for
the combined single lepton, dilepton, and trilepton channels, along with the SSM
$p\bar{p}\rightarrow W'\rightarrow WZ$~production cross section are shown in
Figure~\ref{fig:plot3}. The excluded cross section using only single lepton events and the RS
graviton production cross section are shown in Fig.~\ref{fig:plot4}.

In this analysis we assume a linear relationship between the resonance mass and
total width, and that the width is smaller than the expected
experimental mass resolution. In some classes of models, the total width grows as
a power of the mass yielding widths larger than the expected mass
resolution. Using {\sc pythia}~$W'\rightarrow WZ$~MC events with varying
$W'$~boson widths, we observe that our results are valid for widths below
$10\%$~of the resonance mass or, alternatively, for a coupling strength at the
$W'WZ$~vertex up to ten times the SSM value.

%%%%%%%%%%%%%%%%%%
% CONCLUSION
%%%%%%%%%%%%%%%%%%
In summary, with up to $5.4$~fb$^{-1}$~of Tevatron Run~II integrated luminosity we do not observe
an excess of events over the SM background prediction for events in \ww~and
\wz~boson final states. We set limits on the production cross section multiplied
by the branching ratio for resonant \ww~and \wz~boson 
pair production using two theoretical benchmark scenarios: SSM $W'\rightarrow WZ$ and RS
\gww~production. Under these assumptions, we exclude an SSM~$W'$~boson with a mass
between 180 and 690~GeV and an RS graviton with a mass between 300 and 754 GeV
at $95\%$~CL. Our novel use of the jet mass to select hadronic decays of the
$W$~and $Z$~bosons was essential to obtaining such stringent limits, which are
the best for these new physics scenarios.

% acknowledgement.tex                             6 October 2010
%
We thank the staffs at Fermilab and collaborating institutions,
and acknowledge support from the
DOE and NSF (USA);
CEA and CNRS/IN2P3 (France);
FASI, Rosatom and RFBR (Russia);
CNPq, FAPERJ, FAPESP and FUNDUNESP (Brazil);
DAE and DST (India);
Colciencias (Colombia);
CONACyT (Mexico);
KRF and KOSEF (Korea);
CONICET and UBACyT (Argentina);
FOM (The Netherlands);
STFC and the Royal Society (United Kingdom);
MSMT and GACR (Czech Republic);
CRC Program and NSERC (Canada);
BMBF and DFG (Germany);
SFI (Ireland);
The Swedish Research Council (Sweden);
and
CAS and CNSF (China).


\begin{thebibliography}{99}
\bibitem{deltar} The D0 detector utilizes a right-handed coordinate system with the $z$ axis
pointing in the direction of the proton beam and the $y$ axis pointing upwards. The azimuthal angle 
$\phi$ is defined in the $xy$ plane measured from the $x$ axis. The pseudorapidity is defined as
$\eta = -\rm{ln}[\rm{tan}(\theta/2)]$, where $\theta = \arctan(\sqrt{x^2+y^2}/z)$. The transverse variables are 
defined as projections of the variables onto the $xy$ plane.
\bibitem{metdef} The missing transverse energy, denoted as $\METnoSpace$~in the text, is the
imbalance of the momentum estimated from the calorimeter and reconstructed muons in the $xy$ plane.
\bibitem{mtwz} The transverse mass of a particle with $N$~decay products is defined as \mbox{$M_{T}=\sqrt{(\sum^{N}_{i}(E^{i}_{T})^{2}-(p^{i}_{x})^{2}-(p^{i}_{y})^{2}}$}
\end{thebibliography}
\end{document}